\documentclass[pra,twocolumn,amsmath,amssymb]{revtex4}

\usepackage{graphicx,epsfig}

\newcommand{\ket}[1]{|#1\rangle} 

\def\be{\begin{equation}}
\def\ee{\end{equation}}
\def\bea{\begin{eqnarray}}
\def\eea{\end{eqnarray}}
\def\bma{\begin{mathletters}}
\def\ema{\end{mathletters}}

\def\0{\overline{0}}

\def\q0{\underline{0}}

\def\one{\leavevmode\hbox{\small1\normalsize\kern-.33em1}}

 \def\ket#1{|#1\rangle}

\def\vecb#1{\mathbf{#1}}
\def\ew#1{\langle#1\rangle}

\begin{document}

\title{Quantum Non-Demolition Detection of Strongly Correlated Systems}
\author{Kai Eckert$^1$, Oriol Romero-Isart$^1$,
Mirta Rodriguez$^2$, Maciej Lewenstein$^{2,3}$, Eugene S.
Polzik$^4$, and Anna Sanpera$^{*1,3}$ } \affiliation{$^1$Grup de
F\'isica Te{\`o}rica, Universitat Aut\`onoma de Barcelona,
E-08193, Bellaterra, Spain} \affiliation{$^2$ICFO--Institut de
Ci\`encies Fot\`oniques, E-08860, Castelldefels, Spain}
\affiliation{$^3$ICREA-- Instituci\'o Catalana  de Recerca i
Estudis Avan\c cats, E-08010, Barcelona, Spain}
\affiliation{$^4$Niels Bohr Institute, Danish National Research
Foundation Center QUANTOP, Copenhagen University, Copenhagen 2100,
Denmark}

\begin{abstract}
Preparation, manipulation, and detection of strongly correlated
states of quantum many body systems are among the most important
goals and challenges of modern physics. Ultracold atoms offer an
unprecedented playground for realization of these goals. Here we
show how strongly correlated states of ultracold atoms can be
detected in a quantum non-demolition scheme, that is,  in the
fundamentally least destructive way permitted by quantum
mechanics. In our method, spatially resolved components of atomic
spins couple to quantum polarization degrees of freedom of light.
In this way quantum correlations of matter are faithfully mapped
on those of light; the latter can then be efficiently measured
using homodyne detection. We illustrate the power of such
spatially resolved quantum noise limited polarization measurement
by applying it to detect various standard and "exotic" types of
antiferromagnetic order in lattice systems and by indicating the
feasibility of detection of superfluid order in Fermi
liquids.\end{abstract}


\maketitle

\noindent{\bf Introduction} Future applications of quantum physics
for quantum simulations, computation, communication, and metrology
will require an extremely high degree of control of preparation,
manipulation, and, last but not least, detection of strongly
correlated states of quantum many body systems. Ultracold atoms
offer an unprecedented playground for realization of these goals.
Several paradigm examples of strongly correlated states have been
successfully realized, such as the Mott insulator, the Tonks gas,
and the Bose glass (for a review cf. \cite{review}). A standard
way of analyzing such systems is by releasing the atoms from the
trap and performing destructive absorption spectroscopy, which
only allows to measure the column density of the expanded cloud.
Considerable attention has been thus devoted recently to novel
methods of detection, that allow for measuring (spin)
density-density and other higher order correlation functions. One
of those methods is atomic noise interferometry \cite{demler-det},
whose power is well illustrated in the recent observation of the
bosonic and the fermionic Hanbury Brown-Twiss effect
\cite{Bloch-HB,Bloch-HB2}. Direct atom counting is another way to
measure this effect, and to even go beyond it
\cite{Lewenstein-news}; it can be realized directly with
metastable Helium atoms \cite{alain,alain2}, or by using methods
of cavity quantum electrodynamics (QED)\cite{esslinger}. Cavity
QED is also essential in the recent proposals of Ref.
\cite{ritsch, Murch}, while Ref. \cite{eugene1} proposes how to
prepare and detect magnetic quantum phases using superlattices.
All of the above approaches are, at least in some respects,
destructive and frequently suffer from undesired atom number
fluctuations inevitable in the preparation of the quantum states.

Here we propose a unique method that allows for spatially resolved
quantum non-demolition (QND) \cite{braginsky} detection of quantum
states of ultracold atoms with internal (pseudo-)spin degrees of
freedom. Our approach is based on the idea of  quantum noise
limited polarization spectroscopy, demonstrated in
Ref.~\cite{Sorensen}. Polarization spectroscopy itself consists in
shining a polarized probe laser beam through the atomic spin
system, and has been a subject of studies for years
\cite{massimo}. What is novel in quantum noise limited schemes is
that quantum polarization degrees of freedom of light couple to
the atomic spins, and in this way {\it quantum fluctuations} of
the spins (magnetization) are faithfully mapped onto those of
light; they can then be efficiently measured using homodyne
detection of the transmitted probe. With an appropriate choice of
parameters, and provided that shot-noise-limited detection of
light is achieved, this approach has proved very successful for
QND-based quantum interfaces between light and atoms. In
particular, atomic squeezing \cite{kuzmich}, atomic entanglement,
quantum memory, and teleportation have been achieved in this
context (for a review see \cite{Julsgaard} and references
therein). We have recently proposed to apply this approach to
detect magnetic order of weakly correlated ultracold atoms
\cite{ourPRL}. Unfortunately, such proposal does not provide
spatial resolution, and thus it cannot, e.g., discriminate between
different antiferromagnetic quantum phases present in lattice
models. Here we show that this important limitation can be
overcome by using a standing wave probe laser configuration. Such
a modification is crucial: it allows for spatially periodic QND
coupling of light to atomic spins, and thus can reveal spin
correlations with the period of coupling. The quantum noise of the
transmitted light then carries information on the Fourier
components of the spin density. Controlling the parameters of the
optical lattice and of the probe light standing wave permits
discrimination and characterization of various standard and
"exotic" magnetic orderings.

\begin{figure}[p]
\begin{center}
 \includegraphics[width=0.23\textwidth]{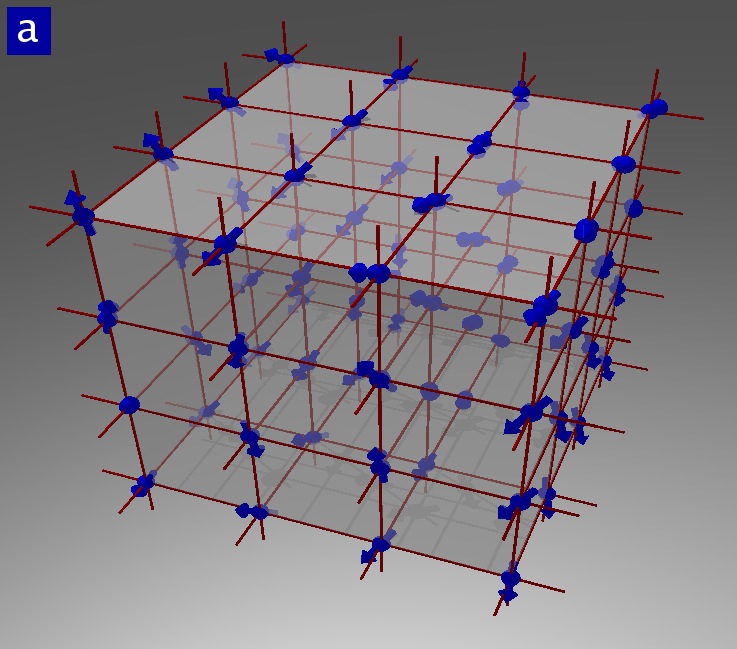}\,\,\,\,
 \includegraphics[width=0.23\textwidth]{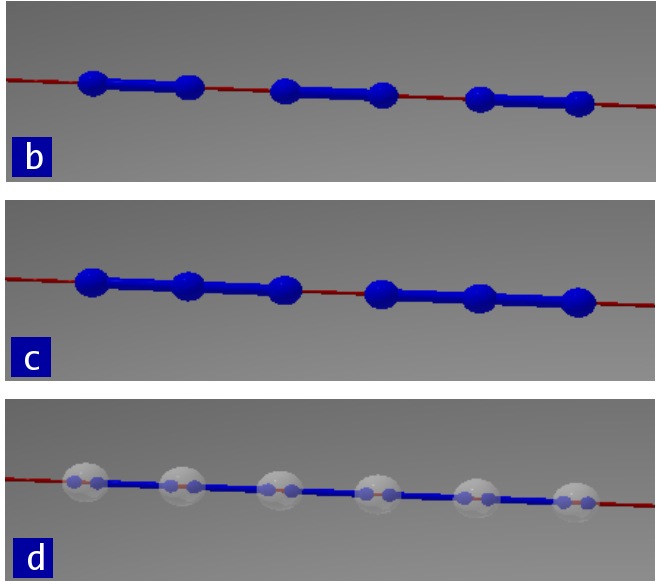}
\caption{{\bf Antiferromagnetic states of spin-$1$ lattice systems}. (a) 3D cubic lattice
with a paramagnetic state of unpolarized atoms; (b) dimerized state with pairs of neighbouring atoms
forming singlets; (c) trimerized state with triples of neighbouring atoms forming singlets;
(d) AKLT (Affleck-Kennedy-Lieb-Tasaky) state obtained from a concatenation of spin-$1/2$
singlets by projecting pairs of spins from different bonds into the subspace of total spin-$1$
\cite{Affleck}.}
\label{fig1}
\end{center}
\end{figure}

In Fig.~\ref{fig1} we schematically depict  various important lattice states, all of
them having zero average of any component of the total spin. The
paramagnetic state, Fig.~\ref{fig1}(a), exhibits large
(proportional to the total number of atoms, $N_{\rm at}$)
fluctuations of the total spin, whereas Fig.~\ref{fig1}(b-d) show
{\it global singlets}, i.e. states with total spin zero, which do
not show any fluctuations. Obviously, those most interesting
strongly correlated {\it global singlet} states of
Fig.~\ref{fig1}(b-d) cannot be distinguished from each other by
spatially homogeneous probing. Probing, however, only every second
atom in, e.g., the dimerized state Fig.~\ref{fig1}(b), i.e.,
measuring the {\it total staggered magnetization} of the state, will result in
its correlations imprinted on the light. As we shall demonstrate
below, such type of measurement offers the possibility to
distinguish between various {\it global singlets}, with minimal
disturbance to the system.

\begin{figure}[p]
\begin{center}
\includegraphics[width=0.47\textwidth]{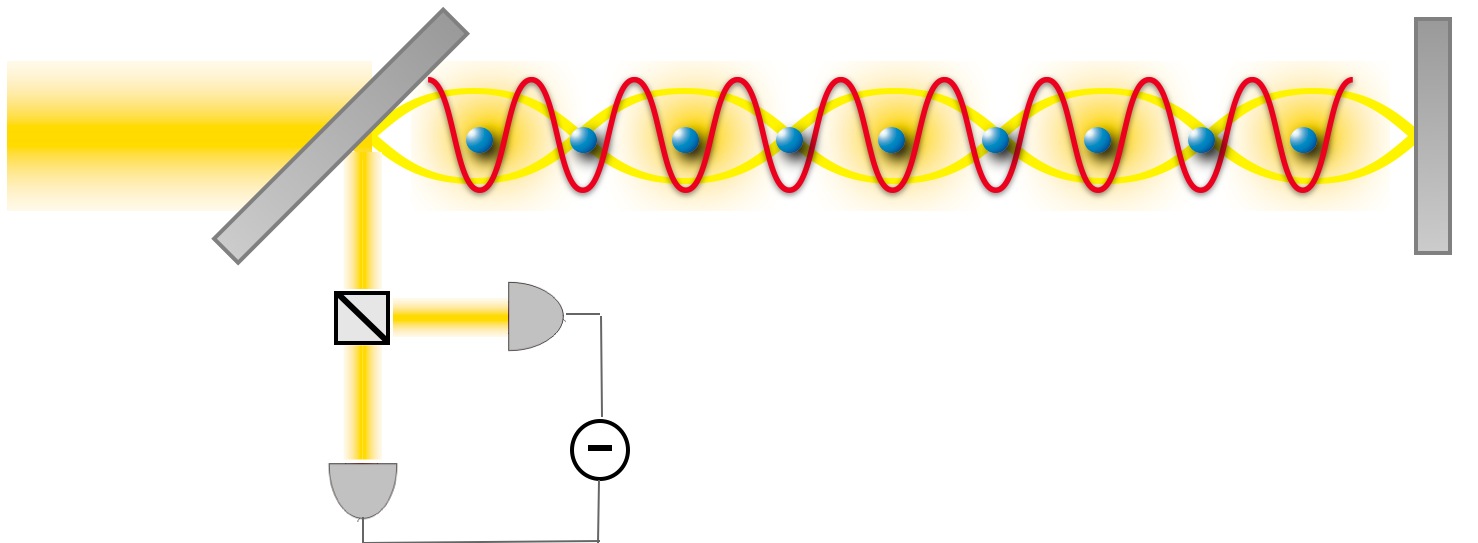}\,\,\,\
\caption{{\bf Schematic experimental setup}. A strong laser beam initially polarized in
$x$-direction is impinging on a $99/1$ beamsplitter. The
transmitted attenuated part of this probe is propagating through
the sample and reflected off a mirror, such that a standing
wave with wavevector $k_P$ in $z$ direction is formed at the
position of the atomic sample. After the second pass,
the laser beam is outcoupled to a homodyne detector,
where $\ew{\hat S_2}$ is recorded. The atoms are trapped in an
optical lattice with wavevector $k$. Its relative displacement
with respect to the probe standing wave can be changed by either moving
the end mirror, or by introducing a phase shift between the
counterpropagating laser beams.}\label{fig:setup}
\end{center}
\end{figure}

\noindent{\bf Detection scheme} The detection mechanism proposed
here uses the off-resonant interaction of spin-$F$ atoms (i.e.,
atoms having a $2F+1$ dimensional ground state manifold) with a
polarized light beam propagating in $z$-direction (for a
theoretical description of light-atom coupling cf.
\cite{Julsgaard,at2}). The light pulse is characterized by
the Stokes operators $\hat s_{1}$, $\hat s_2$, $\hat s_3$
corresponding to the difference of the number of photons in the
$x$ and $y$ linear polarizations, in the $\pm45^\circ$ linear
polarizations, and in the two circular polarizations,
respectively. Let us first consider a one-dimensional (1D) sample
of $N_{\rm at}$ atoms trapped in a 1D optical lattice of period
$\pi/k$, oriented in $z$ direction, with the $n$-th atom located
at position $z_n=(n-1)\pi/k$, having the spin operator
$\hat{\vecb{j}}(z_n)$. The relevant dispersive part of the
interaction Hamiltonian between light and atoms reads
\cite{Julsgaard,at2} (see methods)
\be \hat H=-\kappa \hat
s_{3}\hat{J}^{\rm\ eff}_z,\label{eqn:heff} \ee
 Here $\kappa$ is
the coupling constant and $\hat{J}^{\rm\ eff}_z$ is the
$z$-component of the effective collective atomic spin
$\hat{\vecb{J}}^{\rm\ eff}=\sum_nc_n\hat{\vecb{j}}(z_n)$. The
coefficients $c_n$ account for the modification of the atom-light
coupling due to a spatial modulation of the probe beam intensity
(in a running wave configuration $c_i\equiv 1$) and are the key
parameters allowing for spatial resolution. As a particular case,
we consider a standing wave configuration, as it is described in
the caption of Fig.~\ref{fig:setup}. Then, \be c_n\equiv
c_n(k_P,a)=2\int dz\cos^2(k_P(z-a))|w(z-z_n)|^2,
\ee with $k_P$ as the wavevector of the standing wave probe, and $a$ its shift
with respect to the optical lattice; $w(z-z_n)$ is the usual Wannier-type wave function of a
single atom confined at $z_n$ in a deep optical lattice.

We take the probe beam to be strongly polarized in the $x$ direction so that $\hat S_1$ ($\hat S_i =\int  s_i dt$)
fulfills $\ew{\hat S_1}=N_{\rm ph}/2\gg1$. This fact permits
to introduce canonical quadrature operators $\hat X=\hat
S_2/\sqrt{N_{\rm ph}}$, $\hat P=\hat S_3/\sqrt{N_{\rm
ph}}$, which satisfy $[\hat X, \hat P]\approx\it i$. 
Integrating the Heisenberg equation of motion for $\hat s_{2}$ shows that the
effective collective spin $\hat{J}^{\rm eff}_z$ is imprinted on
the $\hat X$ quadrature of light: \be \hat X_{\rm out}=\hat X_{\rm
in}-\frac{\kappa}{\sqrt{FN_{\rm at}}}\hat{J}^{\rm\
eff}_z.\label{eqn:xout} \ee Since $\ew{\hat X_{\rm in}}=0$, the
mean of this quadrature after passing through the sample is
directly proportional to the mean of the $z$-component of the
effective atomic spin. Its variance contains the shot noise of the
incoming probe pulse and the variance of $\hat{J}^{\rm\ eff}_z$. Assuming
a coherent input beam;
\be \ew{(\Delta \hat X_{\rm
out})^2}=\frac12+\frac{\kappa^2}{FN_{\rm at}}\ew{(\hat{J}^{\rm\
eff}_z- \ew{\hat{J}^{\rm\ eff}_z})^2}. \ee The variance of the
collective atomic spin can be efficiently determined from the
measurement on light if
$\kappa=\sqrt{d\;\eta}\geq1$\cite{Julsgaard}. Here
$d=N_{at}\sigma/A$ denotes the resonant column optical depth of
the atomic sample and $\eta=(N_{ph}\sigma\Gamma^2)/(\Delta^2 A)$
is the probability of resonant excitation per atom by the probe,
where $\sigma$ is the cross section on resonance for the probe
transition, $\Gamma$ the spontaneous decay rate, $\Delta$ the
detuning from resonance and $A$ the cross section of the atomic
ensemble illuminated by the probe. Since spontaneous emission
destroys the spin state and the QND character of the coupling, the
condition $\eta\ll1$ has to be fulfilled.
To evaluate the decoherence (and heating) caused by the probe, we
need to estimate the effect of spontaneous emission on the
variance of the light quadrature. The induced probe decoherence is
given  by $\ew{(\Delta J^{decoh}_{z})^2}/FN_{at}\approx \eta$ (in
units of the coherent spin state noise). Therefore, the total
measured noise reads
 \be \ew{(\Delta \hat X_{\rm
out})^2}=\frac{1}{2}+{\kappa^2}\left[\frac{\ew{(\Delta \hat J_{z}^{\rm
eff})^2}}{FN_{at}}+\eta\right]. \ee
In order to measure the atomic spin
fluctuations with the best possible accuracy,
we need to maximize the {\it signal
to noise ratio}
$[\kappa^2\ew{(\Delta \hat J_{z}^{\rm
eff})^2}/FN_{at}]/[1/2+\kappa^2\eta]$.
For a coherent input beam, this contribution is maximized for
$\eta_{opt}\approx1/\sqrt{2d}$. Notice that $\eta$ can be adjusted
to $\eta_{opt}$, by choosing the appropriate detuning, intensity
and duration of the laser probe. On the other hand, the optical
depth of a sample of ultracold atoms, including e.g. a cubic
lattice with $100\times 100 \times 100$ atoms can easily reach
values of few hundreds. Thus, the value of $\eta$ can be made much
smaller than one \cite{Hammerer2004}. Since the spin noise of
interest (as shown later in Fig. 3) is of the order of unity, the
additional noise provided by spontaneous emission does not,
therefore, significantly modify the measured variance of the light
quadrature.

Furthermore, a squeezing of the $\hat X$ quadrature of the
incoming probe before passing it through the sample such that
$\ew{(\Delta \hat X_{\rm in})^2}<1/2$, allows to improve the
signal-to-noise ratio even further. Finally, let us mention that
from the data recorded at the homodyne detector also higher order
terms $\ew{(\Delta \hat X_{\rm out})^m}$ can be extracted.

In order to achieve a spatial modulation of the atom light
coupling, $k_p/k\neq1$ is necessary. The wavelength of the probe
laser and thus $k_P$ is constrained by the near resonant condition
for an efficient QND probing. Also the choice of the wavelength of
the light forming the optical lattice is limited. Thus, tuning the
wavelengths of the lasers only gives a restricted control over the
ratio $k_p/k$. An appropriate choice of the ratio can, however, be
achieved in various ways: (i) trivially by modifying $k$ by changing the angle
between the beams forming the optical lattice; (ii) more surprisingly, by 
shining the probe light at an angle $\theta$ to the lattice. The effective
Hamiltonian then reads
\be H=-\kappa \hat
s_{3}\left(\cos\theta\hat{J}^{\rm\ eff}_z-\sin\theta\hat{J}^{\rm\
eff}_y\right), \ee
and the effective wavevector for a 1D sample
oriented in $z$-direction is $k_P=2\pi\cos\theta/\lambda_P$, with
$\lambda_P$ the wavelength of the probe laser. Finally, (iii) a
probe standing wave with variable $k_P$ can be obtained from
crossing the probe laser beams at a variable angle. This last
approach leads, however, to a different interacting Hamiltonian.

\begin{figure*}[p]
\begin{center}
\includegraphics[width=0.9\textwidth]{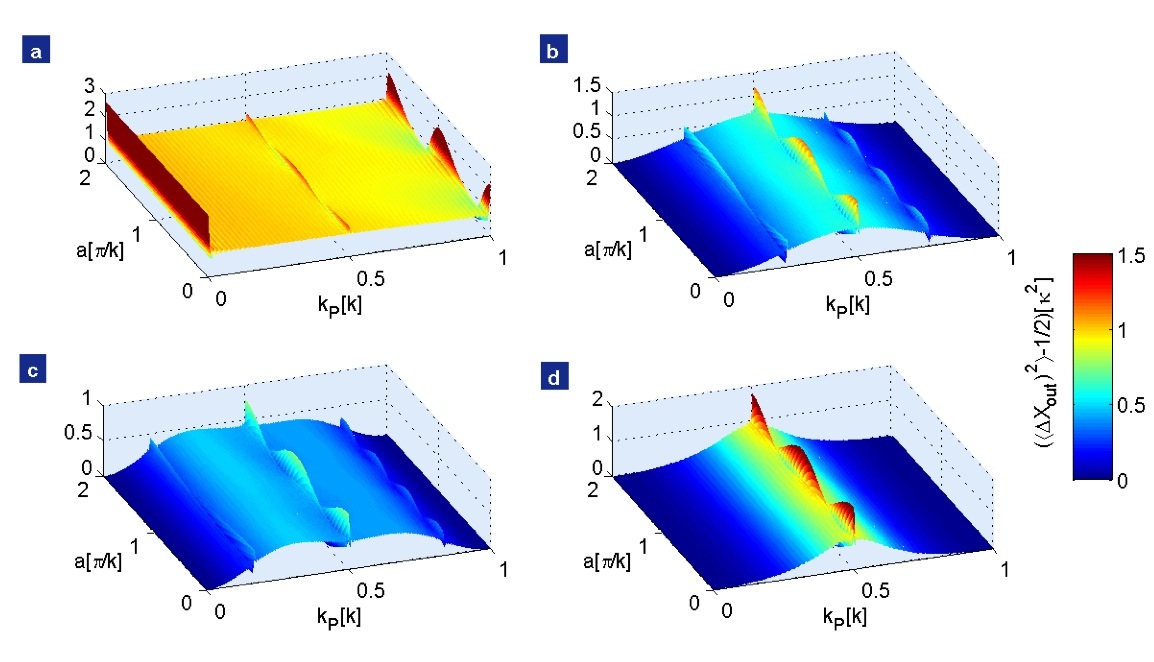}\,\,\,\
\caption{ {\bf Detection of antiferromagnetic states of spin-$1$
lattice systems}. Fluctuations ($\epsilon \equiv \ew{(\Delta \hat X_{\rm out})^2}-1/2$) imprinted on the $\hat X$
quadrature of the probe light beam transmitted through a sample of
spin-$1$ atoms in a one-dimensional lattice for (a) unpolarized
paramagnetic, (b) dimerized, (c) trimerized, and (d) AKLT states.
While the presence of fluctuations at $k_{P}/k=0$ signals the
unpolarized paramagnetic state, the other three global singlets
phases can be unambiguously distinguished at different values of
the ratio $k_{P}/k$ and/or the shift $a$ (see text for
details).}
\label{fig3}
\end{center}
\end{figure*}

\noindent{\bf Quantum antiferromagnets in 1D} Let us now
illustrate the power of our proposal, by applying it to concrete
examples of strongly correlated states that can be realized with
ultracold atoms. Particularly challenging are in this context
various possible quantum antiferromagnetic states that lie in the
center of interest of condensed matter and even high energy
physics \cite{Misguish03,Alet06}. We will analyze some of the
states that appear as ground states (or idealizations thereof) of
the generalized Heisenberg spin-1 atomic chain
\cite{demler,yip,legaza} (see Methods).  For some choice of
parameters such systems are in the, so called, Haldane phase. This
phase is gapped, and the ground state is well described by a
matrix product state (MPS), that is both translationally and
rotationally invariant.   A prominent example of such states is
the AKLT state \cite{Affleck} (see Fig. \ref{fig1}(d)). In another
regime of parameters the ground state exhibits the dimer, or
Peierls, order and it represents an example of the Valence Bond
Solid (VBS) state.  To a good approximation the ground state with
an even number of sites is formed by placing a dimer on every odd
bond (i.e., pairing the atoms on these bonds into singlets, see
Fig. \ref{fig1}(b)). Such a state breaks the translational
invariance, but not the rotational symmetry. Yet for another
choice of parameters, the system is in the critical phase, but
exhibits some trimerization. The ideal trimer state corresponds to
concatenations of triples of neighbouring atoms forming a singlet,
i.e. the state of zero total spin of the triple (Fig.
\ref{fig1}(c)). The result of applying our detection method to the
paramagnetic and these three types of {\it global singlet} states
are shown in Fig.~\ref{fig3}. While the presence of fluctuations
at $k_{P}/k=0$ signals unambiguously the unpolarized paramagnetic
state, the three distinct global singlets can be perfectly
distinguished either by combining results for variable $k_P/k$ and
$a$, or by fixing $k_P/k$ and varying solely $a$. For instance, at
$k_{P}/k=1/2$ and $a=0$, the added noise (in units of the shot
noise of the probe light) for the dimerized, trimerized, and AKLT
states is respectively $4\kappa^2/3$, $8\kappa^2/9$, and
$2\kappa^2$.
The above quoted values correspond the the
ideal case in which the atoms are $\delta$-localized on the minima
of the optical lattice, while Fig. ~\ref{fig3} shows the results
for extended atomic Wannier functions. This fact yields an overall
decrease of the noise contribution towards larger $k_P/k$ ratios.
Except for this decrease, the pattern is repeated for larger
values of $k_P/k$. The fluctuations on the light quadrature using
the atomic ground states of the spin-1 lattice systems obtained
numerically \cite{transportspin1}, present a very close
resemblance to the ones depicted in Fig.~\ref{fig3}.

\begin{figure}[p]
\begin{center}
\includegraphics[width=0.2\textwidth]{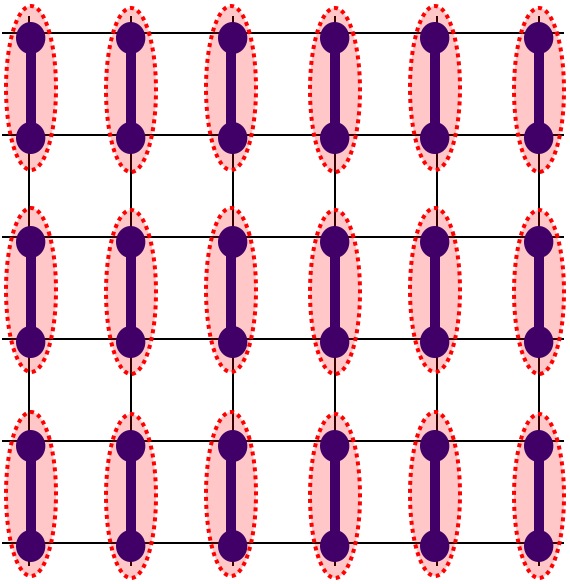}\hspace{2em}
\includegraphics[width=0.2\textwidth]{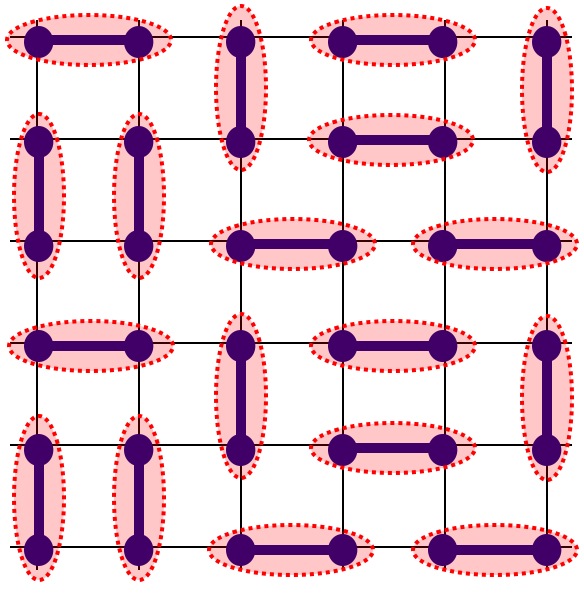}
\caption{
{\bf Two-dimensional states of dimers}. (a) Valence Bond Solid arranged in a regular crystal, (b) Valence Bond
Solid consisting in a random covering of dimers.
}\label{fig4}
\end{center}
\end{figure}

\noindent{\bf Quantum antiferromagnets in 2D} Our scheme is not
limited to arrays of 1D systems. It can also detect and
distinguish different antiferromagnetic ground states  in lattice
models in higher dimensions \cite{Misguish03, Alet06}. As an
illustration, we consider a Heisenberg spin-$1/2$ model on a
square lattice with antiferromagnetic coupling. In this case it
may happen that translational symmetry is broken and the ground
state becomes the dimer VBS state (see Fig.~\ref{fig4}(a)), which
exhibits long range dimer order. Also exotic spin liquid states
which do not break any symmetry and do not possess long range
order can appear in an extended Heisenberg model. These states can
be visualized as coherent superpositions of random dimer
coverings, termed as Resonating Valence Bond (RVB) states (see
Fig. \ref{fig4}(b)). We consider a 2D lattice in the $y-z$ plane,
and the probe light strongly polarized along the $x$-direction
propagating in $z$ direction. Assuming all atoms paired into
dimers, we obtain that the additional noise contribution to the
outgoing $\hat X$ quadrature is proportional to the number of
dimers oriented in the $z$-direction. A valence bond solid with
all dimers oriented in the $z$-direction and a single valence bond
state with randomly oriented dimers can then be easily
distinguished, since for $k_{P}/k=1/2$ and $a=0$ the additional
noise is $2\kappa^2$ and $\kappa^2$, respectively. The
antiferromagnetic N\'eel state appearing in the 2D Heisenberg
model can be indeed detected with the method of Ref.
\cite{ourPRL}.

\noindent{\bf Fermionic superfluidity}
Another application of our scheme is to detect superfluid phases in
fermionic ultracold gases in a QND way.
For the probing standing wave set
in the $z$-direction the transmitted light carries information on the fluctuations of
$\hat{J}_z^{\,\textrm {eff}}= \int d^3r \cos^2(k_P z) \hat{J}_z({\bf r})$.
In a 2D or 3D system one can measure the fluctuations in $\hat J_y$ and
$\hat J_x$ in the $y$ and $x$ direction, respectively, by changing
the standing wave orientation.
Similar quantities have been proposed and proved to
detect superfluidity in ultracold Fermi gases \cite{CFA,CFB,CFC,CFD,CFE}.
Our scheme as described here can be specially interesting for a 1D system as one can
change the probing momentum $k_P$ with respect to the Fermi momentum by
tilting the probing standing wave with respect to the system long axis as discussed above.
Comparison of these fluctuations for a paired and a non paired system
reveals the existence of a pairing gap. Already in a homogeneous set-up one could measure the fluctuations in $\hat J_z$ that, in a similar way as the structure factor,
change dramatically across the BCS to BEC crossover \cite{Littlewood}.

\noindent{\bf Acknowledgements.} We thank J.I. Cirac  for
discussions. We acknowledge support of the EU IP Programmes
SCALA, QAP, COVAQIAL, ESF PESC Programme QUDEDIS,
Spanish MEC grants (FIS 2005-03169/04627, AP2005-0595, EX2005-0830, Consolider-Ingenio2010
CSD2006-00019 QOIT), and Catalan grant SGR-00185. Correspondence and requests for materials should be addressed to A. Sanpera. All authors have contributed equally to this work. \\
\noindent{\bf Competing interests statement}: the authors declare that they have no competing financial interests.

\section*{Methods}

\noindent{\bf Atom-light Hamiltonian} The Hamiltonian of equation (\ref{eqn:heff}) is obtained from the off-resonant coupling to
an atomic dipole transition. The excited atomic states can be
adiabatically eliminated from the dipole-interaction Hamiltonian
under the conditions described in the main text. The dispersive
effects arising from the Stark shift of the atomic levels are
described via an effective Hamiltonian \be \hat H=-\int_0^Ldz\rho
A\left(a_0\hat\phi+a_1\hat{s}_z\hat{j}_z+a_2\left[\hat\phi\hat{j}_z^2
-\hat s_-\hat j_+^2-\hat s_+\hat{j}_-^2\right]\right). \ee Here
$L$ is the length of the atomic sample and $A$ its cross section
overlapping with the probe light propagating in the $z$-direction,
and $\rho\equiv\rho(z)$ is the atomic density. The light is
described via Stokes operators $\hat s_{i}\equiv\hat s_{i}(z,t)$
($\hat s_{\pm}=\hat s_1\pm i\hat s_2$) 
and $\hat\phi$ is the total photon density. The coefficients $a_i$
depend on the laser wavelength and the characteristics of the
atomic transition. The first term, proportional to $a_0$, gives
the (polarization independent) ac Stark shift. In the limit where
the detuning is large compared to the hyperfine splitting of the
excited state $a_2\approx0$, and only the linear QND coupling
between the Stokes operator and atomic spin remains. This is
equivalently written in equation (\ref{eqn:heff}) as a coupling between
the Stokes operator and the effective atomic spin
component $\hat J_z^{\rm eff}$. To proceed, we assume that $\hat
J_z^{\rm eff}$ is time independent. This is valid since: (i) it is
conserved by the Hamiltonian equation (\ref{eqn:heff}), and (ii) the measurement time for a pulsed probe is much shorter than the
atomic spin diffusion time. The outgoing $\hat X$ quadrature of
light is then given by Eq.~(\ref{eqn:xout}) provided that
$a_1^2F^2N_{\rm at}^2/2\ll1$. The coupling constant reads
 $\kappa=a_1\sqrt{N_{\rm at}N_{\rm ph}F/2}$.

\noindent{\bf 1D spin chains}
The strongly correlated states of spin-1 atoms on a 1D lattice
considered in this paper are examples of ground states (or
idealizations thereof) of the generalized spin-1 Heisenberg
Hamiltonian \cite{demler,yip,legaza} \be H=\sum_{\ew{n,n'}}
\cos\beta\,\hat{\vecb{j}}(z_n)\cdot\hat{\vecb{j}}(z_{n'})+\sin\beta\,
\left[\hat{\vecb{j}}(z_n)\cdot\hat{\vecb{j}}(z_{n'})\right]^2. \ee
Due to the interplay between the bilinear and the biquadratic interaction
(parametrized by $\beta$), this Hamiltonian presents three antiferromagnetic
quantum phases: dimerized $\beta \in (-3\pi/4,-\pi/4)$, Haldane $\beta \in (-\pi/4,\pi/4)$, and critical $\beta \in [\pi/4,\pi/2)$. The representative point of
the dimerized phase is found at $\beta=-\pi/2$ where the two-site ground state is a
singlet ({\it dimer}, \cite{demler,yip}):
$\ket{s}_{1,2}=(\ket{+1}_1\ket{-1}_2+\ket{-1}_1\ket{+1}_2-\ket{0}_1\ket{0}_2)/\sqrt3$.
Here we denote by $\ket{m_z=\pm1,0}_n$ the eigenstates of $\hat j_z(z_n)$.
This ground state is analytical only in the thermodynamic limit or for periodic boundary conditions.
In a finite chain with an even number of sites, however, the ground state is close to a concatenation of spin-1 singlets
(see Fig.~\ref{fig1}(b)).  The representative state of the Haldane phase is found at
$\beta=\arctan(1/3)$, where the exact ground state corresponds to the AKLT state (see Fig.~\ref{fig1}(d)).
The critical phase is less well understood, though known to have period three correlations.
At $\beta=\pi/4$ the three-site ground state is a singlet ({\it trimer}), and a caricature of the ground
state for a larger number of sites is given by a concatenation of trimers (cf.~Fig.~\ref{fig1}(c)).
These three isotropic ground states are eigenstates of any component of the total
spin. Thus for $k_P=k$ where all atoms couple equally to the light, no additional noise is
imprinted on the $\hat X$ quadrature. For $k_P\neq k$, in general these states are not
eigenstates of $\hat{J}^{\rm\ eff}_z$ (although  $\langle \hat{J}^{\rm\ eff}_z \rangle=0$) and as we demonstrate
in Fig.~\ref{fig1}(c), the fluctuations permit to discriminate them unambiguously.

\end{document}